# Affinity Discrimination in B cells in Response to Membrane Antigen Requires Kinetic Proofreading[1]

Philippos K. Tsourkas[*], Subhadip Raychaudhuri[*,†,‡,§,2]
[*]*Dept. of Biomedical Engineering*, [†]*Biophysics Graduate Group*,
[‡]*Graduate Group in Immunology*, [§]*Graduate Group in Applied Mathematics*
*University of California, Davis*

Running title: B cell affinity discrimination requires kinetic proofreading

ABSTRACT

B cells signaling in response to antigen is proportional to antigen affinity, a process known as affinity discrimination. Recent research suggests that B cells can acquire antigen in membrane-bound form on the surface of antigen-presenting cells (APCs), with signaling being initiated within a few seconds of B cell/APC contact. During the earliest stages of B cell/APC contact, B cell receptors (BCRs) on protrusions of the B cell surface bind to antigen on the APC surface and form micro-clusters of 10-100 BCR/Antigen complexes. In this study, we use computational modeling to show that B cell affinity discrimination at the level of BCR-antigen micro-clusters requires a threshold antigen binding time, in a manner similar to kinetic proofreading. We find that if BCR molecules become signaling-capable immediately upon binding antigen, there is a loss in serial engagement due to the increase in bond lifetime as $k_{off}$ decreases. This results in decreasing signaling strength as affinity increases. A threshold time for antigen to stay bound to BCR before the latter becomes signaling-capable favors high affinity BCR-antigen bonds, as these long-lived bonds can better fulfill the threshold time requirement than low-affinity bonds. A threshold antigen binding time of ~10 seconds results in monotonically increasing signaling with affinity, replicating the affinity discrimination pattern observed in B cell activation experiments. This time matches well (within order of magnitude) with the experimentally observed time (~ 20 seconds) required for the BCR signaling domains to undergo antigen and lipid raft-mediated conformational changes that lead to association with Syk.

# INTRODUCTION

The strength of B cell signaling in response to stimulation by antigen is known to increase with the affinity of the B cell receptor (BCR) for antigen (Ag), a phenomenon known as affinity discrimination (1-8). B cell affinity discrimination has been observed starting from early membrane-proximal tyrosine phosphorylation events to late events such as lymphokine gene transcription (3). The precise mechanisms by which B cells sense antigen affinity are still the subject of current investigations (9). While the first studies of B cell affinity discrimination focused on antigen encountered in soluble form, recent research shows that antigen fragments presented on the surface of antigen presenting cells (APC) are potent stimulators of B cells (4,9-20).

Further studies show that during contact between B cells and antigen presenting cells, B cells initially encounter antigen through BCRs located on protrusions of the B cell surface (21), resulting in the formation of micro-clusters of BCR/Antigen complexes (8,12,21,22). These micro-clusters are thought to be signaling-active (8,12,21,22), as they trigger an affinity-dependent spreading of the B cell surface over the antigen presenting cell surface, increasing the cell-cell contact area (8). This spreading response leads to further micro-cluster formation at the leading edges (8,21) and is concomitant with B cell synapse formation (8,12). It has also been shown that early signaling events (~ 100 seconds) such as $Ca^{2+}$ flux, as well as antigen accumulation in the immunological synapse, all increase with antigen affinity (8).

However, very little is known about how B cells use their membrane-proximal signaling mechanism to discriminate between membrane antigens of varying affinity at the level of BCR-antigen micro-clusters. In this work, we use an *in silico* computational model of B cell signaling to explore whether kinetic proofreading between BCR and antigen can explain B cell affinity discrimination. Originally proposed as a mechanism for how T cells discriminate between high and low affinity ligands (23), the idea behind kinetic proofreading is that a receptor needs to undergo a series of physical modifications induced by ligand binding in order to become signaling-capable. However, the receptor quickly reverts to its unmodified state if the ligand detaches before the fully modified state is reached. This has the net effect of setting a threshold time that the ligand needs to be bound to a receptor before the latter can become signaling-active (24,25).

Although it bears superficial similarities to the T cell receptor (TCR) and Fc epsilon receptor (FcεRI) signaling systems, the BCR system differs in significant ways from both, and what holds true for them may not be assumed to hold true for the BCR system. In T cells, studies indicate that TCR signaling is a non-monotonic function of antigen affinity, starting from $K_A=10^6$ $M^{-1}$, reaching a peak at $K_A=10^7$ $M^{-1}$, and decreasing thereafter (typically reaching up to $K_A=10^8$ $M^{-1}$) (25,26). If it were possible to extend such studies to affinity values above $K_A=10^8$ $M^{-1}$, the signaling response would show an even stronger decrease with affinity. In contrast, the B cell signaling response increases continually starting from $K_A=10^6$ $M^{-1}$, reaching a ceiling around $K_A=10^{10}$ $M^{-1}$ (2). Importantly, the affinity range over which B cells recognize antigen spans five orders of magnitude ($K_A=10^6$ -$10^{10}$ $M^{-1}$) (2,7,8), a much wider range than T cells ($K_A=10^6$ -$10^8$ $M^{-1}$) (27). For high affinity antigens, a very low dissociation rate ($k_{off}$) makes it difficult for an antigen to serially trigger multiple B cell receptors. Furthermore, BCR is a bivalent molecule, whereas both TCR and FcεRI are monovalent, and BCR is moreover expressed at much higher concentrations than TCR. This has the net effect of greatly increasing the avidity of the BCR system as compared to the TCR and FcεRI systems. The question of how a B cell can discriminate between high affinity antigens is thus highly non-trivial and cannot be addressed by extrapolating what is known from TCR and FcεRI studies.

We show that a monotonically increasing B cell response at the level of BCR-antigen micro-clusters requires that antigen be bound to a BCR molecule for a threshold time of ~10 seconds before that BCR's signaling domains become signaling-active, in a manner similar to kinetic proofreading. It has been hypothesized that BCR undergoes conformational changes induced by antigen ligation and interactions with lipid rafts before engaging the Src-family kinase Lyn (21). These conformational changes lead to an opening up of the BCR ITAMs that allows phosphorylation of Syk, but occur within a finite time after the initiation of antigen binding (21,22,28,29).

# METHOD

We investigate B cell affinity discrimination by means of successive *in silico* virtual experiments. Our technique is a Monte Carlo simulation method that builds on our previous work and has been extended to include membrane-proximal signaling events in addition to receptor-antigen binding (30-32). Individual BCR and antigen molecules are explicitly simulated as discrete agents diffusing on virtual cell surfaces and reacting with each other subject to probabilistic parameters that directly correspond to kinetic rate constants.

*Simulation Setup*

Because we are interested in the early stages of antigen recognition, we model a single protrusion on a B cell surface, its cytoplasmic interior, and its vertical projection onto a planar bilayer surface containing antigen, in order to simulate B cell activation experiments as closely as possible (7-9). Lyn is anchored to the B cell protrusion surface, and Syk is uniformly distributed at random in the protrusion's cytoplasm. At the start of a simulation run, all molecules are distributed uniformly at random over their respective domains. At each time step, molecules in the population are individually sampled in a random manner to undergo either diffusion or reaction, determined by means of a coin toss with probability 0.5 (30-32).

*Reaction*

If a molecule has been selected to undergo reaction, we check the corresponding node on the apposing surface for a binding partner. If that is the case, a random number trial with probability $p_{on(i)}$ is performed to determine if the two molecules will form a bond. BCR molecules are bivalent and can bind up to two monovalent antigen molecules, one on each Fab domain. The probability of BCR-antigen binding is denoted by $p_{on(BA)}$. If a BCR/Ag complex is selected, the antigen may dissociate with probability $p_{off(BA)}$ if subsequently sampled to undergo reaction. The reaction probabilities $p_{on}$ and $p_{off}$ are directly analogous to the kinetic rate constants $k_{on}$ and $k_{off}$, and their ratio, denoted as $P_A$, directly analogous to affinity $K_A$. Anchored Lyn can bind to either Ig-α or Ig-β with probability $p_{on(Lyn)}$ and dissociate with probability $p_{off(Lyn)}$. We introduce a threshold antigen binding time $\mu$ such that Lyn can only bind to the Ig-α or Ig-β subunits of a BCR molecule that has bound the same antigen molecule for a length of time $\mu$. Once a BCR has bound antigen for time $\mu$, the BCR remains signaling-capable for the duration of the simulation, even if the antigen subsequently detaches. The length of the threshold time $\mu$ is varied in our simulations. Lyn that is attached to either Ig-α or Ig-β may phosphorylate the Ig-α and Ig-β with probability $p_{phos(Ig\alpha)}$ and $p_{phos(Ig\beta)}$, respectively. Two random number trials, one with probability $p_{phos(Ig\alpha)}$ and the other with probability $p_{phos(Ig\beta)}$ are conducted every time an Ig-α or Ig-β subunit with Lyn attached to it is selected to undergo reaction. Syk can bind to phosphorylated Ig-α or Ig-β with probability $p_{on(Syk)}$ and detach which probability $p_{off(Syk)}$. A Syk molecule that is attached to phosphorylated Ig-α or Ig-β may in turn become phosphorylated with probability $p_{phos(Syk)}$. The phosphorylation trial is carried out every time an Ig-α or Ig-β with an attached Syk molecule is selected for reaction. A schematic of our simplified model of membrane-proximal B cell signaling is shown in Figure 1.

There are a total of 30 possible reactions (all reversible, and not including phosphorylation reactions) and 18 possible species (e.g. free BCR, free Ag, BCR/Ag, BCR/Ag/Lyn, BCR/Ag$_2$ BCR/Ag$_2$/Lyn, BCR/Ag$_2$/Lyn/Lyn, BCR/Ag$_2$/Lyn/Syk, etc…, not including phosphorylation). For BCR-antigen binding, $p_{on}$ and $p_{off}$ vary with the local vertical separation between the B cell surface and the bilayer, $z$, in accordance with the linear spring model (30,31,33,34), while they are uniform for Lyn and Syk binding to Ig-α or Ig-β.

*Diffusion*

If a molecule has been selected to undergo diffusion, a random number trial with probability $p_{diff(i)}$ is used to determine whether the diffusion move will occur. The diffusion probability $p_{diff}$ is directly analogous to the diffusion coefficient $D$. The probability of diffusion of free molecules is denoted by $p_{diff(F)}$, and that of receptor-ligand complexes by $p_{diff(C)}$. If the trial with probability $p_{diff(i)}$ is successful, a direction is selected at random (4 possibilities for surface species, 6 for Syk) and the appropriate neighboring nodes in that direction are checked for occupancy. Molecules may only diffuse if all the required neighboring nodes are vacant, as no two molecules are allowed to occupy the same node. The spacing between nodes is set to 10 nm. Complexes are generally assumed to diffuse slower than free molecules (21), hence $p_{diff(C)}$ is an order of magnitude lower than $p_{diff(F)}$.

*Sampling and time step size*

In our algorithm, the entire molecular population is randomly sampled $M$ times for diffusion or reaction during every time step. The number of trials $M$ is set equal to the total number of molecules (free plus complex) present in the system at the beginning of each time step, and the simulation is run for a number of time steps $T$. A distinguishing feature of our method is a mapping between the probabilistic parameters of the Monte Carlo simulation and their physical counterparts. We do this by setting $p_{diff}$ of the fastest diffusing species in our simulation equal to 1 and matching that quantity to that species' diffusion coefficient $D$. Since the nodal spacing is fixed and known, this allows us calculate the physical length of time that one of our simulation's time steps. Once the time step size is known, it is possible to map $p_{on}$, $p_{off}$, and their ratio $P_A$ to their respective physical counterparts, $k_{on}$, $k_{off}$, and $K_A$. A detailed description of the mapping process can be found in our previous work (30). Such a mapping makes it possible to compare our model's results to those of physical experiments to within an order of magnitude without *a priori* knowledge of the simulation time scale.

*Model parameters*

The parameters used in our model are listed in Table I. Parameter values found in the literature are given on the left side of Table I, while the appropriately mapped forms used in our model are listed on the right side of Table I. BCR-antigen affinity is varied exactly as in B cell activation experiments, by keeping $k_{on}$ constant and varying $k_{off}$, (7,8). Parameters whose values vary during experiments (such as BCR-antigen affinity and antigen concentration) are also varied in our simulations. The same applies for parameters for which we were not able to find measured values in the literature, such as the number of Lyn and Syk molecules ($L_0$, $S_0$), $p_{on(Lyn)}$, $p_{off(Lyn)}$, $p_{on(Syk)}$, $p_{off(Syk)}$, $p_{phos(Igα)}$, $p_{phos(Igβ)}$, and $p_{phos(Syk)}$. For the purposes of obtaining ballpark values for these parameters, we have adapted the values used in modeling studies of FcεRI-mediated signaling that bears many similarities to BCR-mediated signaling (35,36). We have been able to find values for the $K_A$ of Syk binding to Ig-α or Ig-β (37), and hence the ratio $p_{on(Syk)}/p_{off(Syk)}$ is kept fixed in our simulations. The parametric studies conducted to gauge the effect of parameters that are varied in our simulations are included as Supplemental Data.

# RESULTS

*Histogram plots of the number of bound antigens show affinity discrimination as $k_{off}$ decreases*

We investigate affinity discrimination by tabulating the number of bound antigen molecules, the number of B cell receptors with one or more phosphorylated ITAMs (denoted as pBCR), and the number of activated Syk molecules (denoted as aSyk) at the end of a simulation run of 100 physical seconds (i.e. $10^5$ time steps). Because our simulation is stochastic in nature, the number bound antigen, pBCR, and aSyk molecules will vary from one simulation run to another. Each run of our simulation can be thought of as an *in silico* virtual experiment involving a single B cell protrusion. Thus, we perform one hundred independent trials and plot histograms of the results. In Figure 2, we plot the number of bound antigen molecules as BCR-antigen affinity is varied by an order of magnitude across the physiological range, $K_A=10^5$ M$^{-1}$ to $K_A=10^{10}$ M$^{-1}$, as is done in B cell affinity discrimination experiments (7,8). As expected, the number of bound antigen molecules increases with BCR-antigen affinity.

*Histogram plots show affinity discrimination requires a threshold time of antigen binding*

In Figure 3, we plot histograms of the number of pBCR and aSyk molecules for threshold time values of $\mu=0$, 1 and 10 seconds. In the case of pBCR, we observe that when the threshold time $\mu=0$, i.e. BCR becomes signaling-capable immediately upon binding antigen, the histogram plots move in the decreasing direction as affinity increases, indicating weaker signaling with increasing affinity. This is exactly the opposite of what B cell affinity discrimination experiments show (8). With a threshold time of $\mu=1$ second, the histograms are overlapping, with the exception of the histogram for the lowest BCR-antigen affinity value, $K_A=10^5$ M$^{-1}$. Thus it only is possible to distinguish between this affinity value and the rest. This result shows that a threshold time of 1 second is insufficient to induce affinity discrimination in B cells except between the two lowest affinity values. In addition, the histogram for the highest affinity value, $K_A=10^{10}$ M$^{-1}$, shows the maximum number of pBCR and aSyk does not occur at this affinity value, indicating non-monotonic dependence of signaling strength on affinity. When the threshold time is set to $\mu=10$ seconds, the histograms are well separated and show a monotonic increase with affinity. In this instance, it is possible to easily distinguish between all but the two highest affinity values, $K_A=10^9$ M$^{-1}$ and $K_A=10^{10}$ M$^{-1}$, while the number of pBCR is zero for every trial for BCR-antigen affinity $K_A=10^5$ M$^{-1}$. Increasing the threshold time past 10 seconds results in a loss in affinity discrimination at the lower end of the affinity spectrum. Thus, a threshold time of ten seconds is optimal for affinity discrimination at the level of phosphorylated BCR ITAMs. This finding correlates well (within the same order of magnitude) with recent FRET experiments that show that the Ig-α/β signaling subdomains undergo conformational changes that allow interaction with Syk approximately 20 seconds after the initiation of antigen binding (21,28).

In the case of activated Syk molecules, when the threshold time is $\mu=0$, the histograms are overlapping and it is impossible to distinguish affinity values, although perhaps a weakly decreasing trend can be discerned. For a threshold time of $\mu=1$ second, it only is possible to distinguish between $K_A=10^5$ M$^{-1}$ and higher affinity values. For a threshold time of $\mu=10$ seconds, however, the number of aSyk increases with affinity and it is possible to easily distinguish between all but the two highest affinity values. It thus appears that a threshold time of ten seconds is also optimal for affinity discrimination at the level of activated Syk molecules.

Of particular interest is that the number of pBCR and aSyk molecules is zero for the lowest affinity value, $K_A=10^5$ M$^{-1}$, when $\mu=10$ seconds. This replicates the threshold of B cell activation of $K_A=10^6$ M$^{-1}$ seen in experiments (2,4,7). The difficulty in differentiating between the two highest affinity values, $K_A=10^9$ M$^{-1}$ and $K_A=10^{10}$ M$^{-1}$, is also seen in B cell activation experiments, and indicates the existence of a ceiling in B cell affinity maturation around $K_A=10^{10}$ M$^{-1}$ (2,4,7). The results for $\mu=10$ seconds are thus broadly in agreement with experimental investigations of B cell activation.

*Trial-averaged quantities show affinity discrimination requires a threshold time of antigen binding*

In addition to histograms of the number of bound antigen, pBCR and aSyk molecules, we also plot the trial-averaged value of these quantities in Figure 4. Trial-averaged quantities are important as they represent the signaling response integrated from either (a) multiple BCR-antigen micro-clusters within a single cell or (b) from a population of cells. As shown in Figure 4, the trial-averaged number of bound antigen increases monotonically with affinity, as expected, and does not vary with the threshold time $\mu$, as the threshold time only affects events downstream of antigen binding.

The number of pBCR, by contrast, is highly dependent on threshold time. In Figure 4, we observe that when the threshold time $\mu$ is zero, the trial-averaged number of pBCR decreases monotonically with increasing affinity. This is because in our simulations, as in experiments (7), affinity is increased by decreasing the dissociation probability $p_{off}$ (analogous to the dissociation rate $k_{off}$). Higher affinity thus means lower $p_{off}$ and a longer bond lifetime. Long-lived bonds result in fewer encounters between BCR and antigen molecules, as most antigens stay bound to the same BCR molecule for a longer time. Since antigen is the limiting reagent, this means many BCR molecules never encounter antigen. Short-lived bonds, however, result in a rapid succession of binding and unbinding events between BCR and antigen, ensuring most BCR molecules encounter antigen at some point during the simulation. This effect, dubbed "serial triggering" (39,40), is entirely dominant in the case of zero threshold time. For this threshold time, the decrease in serial triggering due to increasing bond lifetime as affinity increases results in fewer signaling-capable BCRs as affinity increases. Thus, downstream signaling can negate the affinity discrimination seen at the level of bound antigen on the surface. By contrast, when the threshold time is set to $\mu=10$ seconds, the number of pBCR increases monotonically with affinity. This shows that kinetic proofreading is dominant at this threshold time value. As Lyn can only phosphorylate BCR molecules that have bound the same antigen molecule for 10 seconds or longer, the shorter bond lifetime associated with low affinity results in few BCR molecules that meet this criterion at low affinity, but many BCR molecules that do so at high affinity. This leads to an increase in the number of phosphorylation events, and hence in the number of pBCR and aSyk molecules, as affinity increases. For the case of a threshold time of $\mu=1$ second, the number of pBCR varies non-monotonically with increasing affinity, indicating a competition between serial triggering and kinetic proofreading. Kinetic proofreading appears dominant at the lower end of the affinity range, while serial triggering appears to dominate at the higher end, with signaling strength reaching its peak at mid-range affinity values. Such a balance between kinetic proofreading and serial triggering leads to the non-monotonic signaling activation in T cells (25,26), but not in B cells.

The pattern in the number of aSyk molecules follows that of pBCR for all threshold time values, as Syk activation occurs downstream of BCR ITAM phosphorylation. Taken together, these results indicate that B cell affinity discrimination requires a kinetic proofreading-type mechanism involving a threshold time greater than one second but no greater than 10 seconds.

*Time course of signaling activation*

In Figure 5, we plot the number of bound antigen as a function of time for affinity values in the range $K_A=10^5$ M$^{-1}$ to $K_A=10^{10}$ M$^{-1}$. The number of bound antigen increases rapidly at first, then slows down as it approaches equilibrium. The change in number of bound antigen with time is not affected by changes in the threshold time $\mu$.

In Figure 6, we plot the time evolution of the number pBCR and aSyk for each order of magnitude in affinity between $K_A=10^5$ M$^{-1}$ to $K_A=10^{10}$ M$^{-1}$. Threshold time $\mu=0$ is shown in the top row, $\mu=1$ second in the middle row, and $\mu=10$ seconds in the bottom row. The patterns observed in the Figures 3 and 4 are observed here as well. For threshold time $\mu=0$ (Fig. 6A,D), the decrease in pBCR and aSyk with increasing affinity seen in Figures 3A and 3D is readily observable. For threshold time $\mu=1$ second (Fig. 6B,E) it only is possible to distinguish between $K_A=10^5$ M$^{-1}$ and the rest, just as in Figures 3B and 3E. For threshold time $\mu=10$ seconds (Fig. 6C,F), the increase in pBCR with increasing affinity is observable, and it is possible to distinguish among affinity values, as in Figure 3C and 2F. The number of pBCR is zero at all time for $K_A=10^5$ M$^{-1}$ at this threshold time.

*Off-rate as the principal regulator of affinity discrimination*

Experimental studies have shown that antigens with similar $K_A$ but different $k_{on}$ and $k_{off}$, can result in very different signaling responses (3,7). We thus also probed the effect of varying $p_{on(BA)}$ and $p_{off(BA)}$ in tandem while keeping the ratio $P_{A(BA)}$ constant. Using $p_{on(BA)}=1.0$ and values of $p_{off(BA)}$ one order of magnitude higher than those in Table I resulted in a predictable increase in the number of BCR-antigen complexes across the board. However, the number of pBCR and aSyk for the two lowest affinity values ($K_A=10^5$ M$^{-1}$, $K_A=10^5$ M$^{-1}$) was zero for $\mu=10$ seconds. This is because $k_{off}$ is the sole determinant of whether an antigen will stay bound to a BCR long enough for that BCR to satisfy the threshold time requirement. The results for $P_{A(BA)}=0.1/0.01$ and $P_{A(BA)}=1/0.01$ (which correspond to $K_A=10^5$ M$^{-1}$ and $K_A=10^5$ M$^{-1}$, respectively) are thus identical as far as the number of pBCR and aSyk is concerned.

## DISCUSSION

In this study, we have shown that affinity discrimination in B cells requires a kinetic-proofreading-type mechanism, in which antigen needs to stay bound to BCR for a threshold time of ~10 seconds before the Ig-α and Ig-β subunits of BCR become signaling-active. If BCR becomes signaling-active before this threshold time of antigen engagement, we fail to observe affinity discrimination as seen in B cell activation experiments.

Experimental studies of B cell activation show a significant change in FRET intensity between BCR cytoplasmic chains within a few seconds of BCR encountering antigen (21,28). This suggests that a lipid-raft mediated conformational change (or a series of conformational changes) occurs in BCR upon encountering membrane antigen. What is intriguing is that the above-mentioned FRET experiments show a sharp increase, followed by a decrease, in intracellular FRET between BCR signaling domains for a time scale of the order of ~ 10-100 seconds (21). Based on this finding, Tolar et al. propose a mechanism of B cell signaling by which B cell receptors undergo a sudden conformational change to a signaling capable "open" conformation after a threshold time following antigen binding. This conformational change could serve as the physical basis of the threshold time proposed in our kinetic proofreading model (21).

We show that if BCR molecules become signaling-capable immediately after binding antigen, the decrease in serial engagement as affinity (and thus bond lifetime) increases results in less BCR ITAM phosphorylation and hence weaker signaling. This is the opposite of what is observed in B cell activation experiments (7,8). Imposing a requirement that antigen be bound to a BCR for a threshold length of time before the Src-family kinase Lyn can bind that BCR's Ig-α and Ig-β signaling subunits, in a manner similar to kinetic proofreading, significantly improves affinity discrimination. In this case, our model reproduces affinity discrimination patterns in which signaling strength increases monotonically with affinity, as observed in experimental investigations of B cell activation (8). We find that affinity discrimination is optimal for a threshold time of ~10 seconds. This time matches well (within order of magnitude) with the experimentally observed time (~ 20 seconds) required for BCR signaling domains to undergo antigen and lipid raft-mediated conformational changes that lead to association with Syk (21,28).

A threshold time of ~1 second results in a pattern in which signaling strength initially increases with affinity at the low end of the affinity range, reaches a maximum value at mid-range affinity, and subsequently decreases. This reflects a balance between kinetic proofreading and serial triggering, and is observed in T cells (25,26), but is not the case in B cells. Also of importance is that our simulation replicates the BCR-antigen affinity threshold of $K_A=10^6$ M$^{-1}$ and ceiling of $K_A=10^{10}$ M$^{-1}$ observed in B cell activation experiments (2,4,7).

A graded signaling response at the level of micro-clusters can then be integrated (from many such clusters) inside B cells into a graded downstream response that will lead to affinity-dependent spreading of the B cell surface (8). This will consequently lead to affinity dependent collection of antigens in the B cell immunological synapses (7,8). Thus, one of the major functions of the B cell immunological synapse could be to collect antigen in an affinity-dependent manner as BCR-antigen affinity increases.

Our model has the distinguishing feature that the probabilistic, dimensionless parameters it employs can be mapped onto their physical counterparts, allowing a meaningful physical interpretation of the results. A threshold time of 1000 dimensionless simulation time steps can thus be mapped into a physical time of 10 seconds, for example. The prediction of a ~10s threshold time is not sensitive to variations in the values of parameters such as the number of antigen, Lyn and Syk molecules, Lyn and Syk on/off rate, or the phosphorylation rate of Ig-α, Ig-β, and Syk (see Supplemental Information). Although our model represents a simplified version of the B cell receptor signaling pathway, it captures the essential details of the early stages of B cell activation and sheds insight into this important immunological process.

## ACKNOWLEDGEMENTS

The authors thank Dr. Emanual Maverakis for proofreading the manuscript.

REFERENCES

1. Lanzavecchia, A. 1985. Antigen-specific interaction between T and B cells. *Nature* 314:537-539.

2. Batista, F.D., and M. Neuberger. 1998. Affinity dependence of the B-cell response to antigen: A threshold, a ceiling, and the importance of off-rate. *Immunity* 8:751-759.

3. Kouskoff, V., *et al.* 1998. Antigens varying in affinity for the B cell receptor induce differential B lymphocyte responses. *J. Exp. Med.* 188:1453-1464.

4. Batista, F.D., and M. Neuberger. 1998. B cells extract and present immobilized antigen: Implications for affinity discrimination. *EMBO J.* 19:513-520.

5. Shih, T.A., E. Meffre, M. Roederer M, and M.C. Nussenzweig MC. 2002a. Role of antigen receptor affinity in T cell-independent antibody responses in vivo. *Nat. Immunol.* 3:399-406.

6. Shih, T.A., E. Meffre, M. Roederer M, and M.C. Nussenzweig MC. 2002b. Role of BCR affinity in T cell dependent responses in vivo. *Nat. Immunol.* 3:570-575.

7. Carrasco, Y.R., S.J. Fleire, T. Cameron, M.L. Dustin, and F.D. Batista. 2004. LFA-1/ICAM-1 interaction lowers the threshold of B cell activation by facilitating B cell adhesion and synapse formation. *Immunity* 20:589-599.

8. Fleire, S.J., *et al.* 2006. B cell ligand discrimination through a spreading and contracting response. *Science* 312:738-741.

9. Carrasco, Y.R., and F.D. Batista. 2006. B cell recognition of membrane-bound antigen: An exquisite way of sensing ligands. *Curr. Op. Immunol.* 18:286-291.

10. Batista, F.D., D. Iber, and M.S. Neuberger. 2001. B cells acquire antigen from target cells After synapse formation. *Nature* 411:489-494.

11. Carrasco, Y.R., and F.D. Batista. 2007. B cells acquire particulate antigen in a macrophage-rich area at the boundary between the follicle and the subcapsular sinus of the lymph node. *Immunity* 27:160–171.

12. Depoil, D., *et al.* 2008. CD19 is essential to B cell activation by promoting B cell receptor-antigen micro-cluster formation in response to membrane-bound ligand. *Nat. Immunol.* 9:63-72.

13. Batista, F.D., and N.E. Harwood. 2009. The who, how, and where of antigen presentation to B cells. *Nat. Rev. Immunol.* 9:15-27.

14. Carrasco, Y.R., and F.D. Batista. 2006. B-cell activation by membrane-bound antigens is facilitated by the interaction of VLA-4 with VCAM-1. *EMBO J.* 25: 889–899.

15. Bergtold, A., D.D. Desai, A. Gavhane, and R. Clynes. 2005. Cell surface recycling of internalized antigen permits dendritic cell priming of B cells. *Immunity* 23:503–514.

16. Junt, T., *et al.* 2007. Subcapsular sinus macrophages in lymph nodes clear lymph-borne viruses and present them to antiviral B cells. *Nature* 450:110–114.

17. Phan, T.G., I. Grigorova, T. Okada, and J.G. Cyster. 2007. Subcapsular encounter and complement-dependent transport of immune complexes by lymph node B cells. *Nat. Immunol.* 8:992-1000.

18. Qi, H., J.G. Egen, A.Y. Huang, and R.N. Germain. 2006. Extrafollicular activation of lymph node B cells by antigen-bearing dendritic cells. *Science* 312:1672–1676.

19. Schwickert, T.A., *et al.* 2007. In vivo imaging of germinal centers reveals a dynamic open structure. *Nature* 446:83–87.

20. Balázs, M., F. Martin, T. Zhou, and J. Kearney. 2002. Blood dendritic cells interact with splenic marginal zone B cells to initiate T-independent immune responses. *Immunity* 17:341–352.

21. Tolar, P., H.W. Sohn, and S.K. Pierce. 2008. Viewing the antigen induced initiation of B cell activation in living cells. *Immunol. Rev.* 221: 64-76.

22. Sohn, H.W., P. Tolar, and S.K. Pierce. 2008. Membrane heterogeneities in the formation of B cell receptor–Lyn kinase microclusters and the immune synapse. *J. Cell Bio.* 182:367-379.

23. McKeithan, T.W. 1995. Kinetic proofreading in T-cell receptor signal-transduction. *Proc. Natl. Acad. Sci. USA* 92:5042–5046.

24. Goldstein, B., J.R. Faeder, and W.S. Hlavacek. 2004. Mathematical and computational models of immune-receptor signaling. *Nat. Rev. Immunol.* 4:445-456.

25. Coombs, D., A.M. Kalergis, S.G. Nathenson, C. Wofsy, and B. Goldstein B. 2002. Activated TCRs remained marked for internalization after dissociation from pMHC. *Nat. Immunol.* 3:926-931.

26. Kalergis, A.M., *et al.* 2001. Efficient T cell activation requires an optimal dwell-time of interaction between the TCR and the pMHC complex. *Nat. Immunol.* 2:229–234.

27. Grakoui, A., *et al.* 1999. The immunological synapse: A molecular machine controlling T cell activation. *Science* 285:221-227.

28. Tolar, P., H.W. Sohn, and S.K. Pierce. 2005. The initiation of antigen induced BCR signaling viewed in living cells by FRET. *Nat. Immunol.* 6:1168 – 1176.

29. Tolar, P., J. Hanna, P. Krueger, and S.K. Pierce. 2009. The constant region of the membrane Immunoglobulin mediates B Cell-Receptor Clustering and Signaling in Response to Membrane Antigens. *Immunity* 30:44-55.

30. Tsourkas, P., N. Baumgarth, S.I. Simon, and S. Raychaudhuri. 2007. Mechanisms of B cell synapse formation predicted by Monte Carlo simulation. *Biophys. J.* 92:4196-4208.

31. Tsourkas, P., M.L. Longo, and S. Raychaudhuri. 2008. Monte Carlo simulation of single molecule diffusion can elucidate the mechanism of B cell synapse formation. *Biophys. J.* 95:1118–1125.

32. Raychaudhuri, S., P. Tsourkas, and E. Willgohs. 2009. Computational modeling of receptor-ligand binding and cellular signaling processes. *In* Handbook of Modern Biophysics, Volume I Fundamentals, T. Jue, editor. Humana Press, Springer, New York.

33. Bell, G.I. 1983. Cell-cell adhesion in the immune system. *Immunol. Today* 4:237-240.

34. Dembo, M., T.C. Torney, K. Saxman, and D. Hammer. 1988. The reaction-limited kinetics of membrane-to-surface adhesion and detachment. *Proc. R. Soc. Lond. B* 234:55-83.

35. Wofsy, C., C. Torigoe, U.M. Kent, H. Metzger, and B. Goldstein. 1997. Exploiting the difference between intrinsic and extrinsic kinases: implications for regulation of signaling by immunoreceptors. *J. Immunol.* 159:5984–5992.

36. Faeder, J.R., *et al*. 2003. Investigation of early events in FcεRI-mediated signaling using a detailed mathematical model. *J. Immunol.* 170:3769–3781.

37. Tsang, E., *et al*. 2008. Molecular mechanism of the Syk activation switch. *J. Biol. Chem.* 283:32650-32659.

38. Favier, B., N.J. Burroughs, L. Weddeburn, and S. Valitutti. 2001. T cell antigen receptor dynamics on the surface of living cells. *Int. Immunol.* 13:1525-1532.

39. Valitutti, S., S. Muller, M. Cella, E. Padovan, and A. Lanzavecchia. 1995. Serial triggering of many T-cell receptors by a few peptide–MHC complexes. *Nature* 375:148–151.

40. Valitutti, S., and A. Lanzavecchia. 1997. Serial triggering of TCRs: A basis for the sensitivity and specificity of antigen recognition. *Immunol. Today* 18:299–304.

[1]S.R. and P.T. are supported by the NIH grant AI074022
[2]Address correspondence to: Subhadip Raychaudhuri, Dept. of Biomedical Engineering, University of California, One Shields Avenue, Davis, CA 95616, Tel: (530) 754-6716
raychaudhuri@ucdavis.edu

TABLE LEGEND

**Table I.** Experimentally measured parameter values found in the literature and the mapped probabilistic counterparts used in our simulations. Ballpark values are used for parameters whose values we were not able to find in the literature, and parametric studies were conducted to gauge their effect on the results.

| Experimental Parameter | Measured or Estimated Value | Simulation Parameter | Mapped Value |
|---|---|---|---|
| $K_A$ BCR-Ag | $10^6$-$10^{10}$ M$^{-1}$ ‡ (7,8) | $P_{A(BA)}$ | $10^2$-$10^6$ |
| $k_{on}$ BCR-Ag | $10^6$ M$^{-1}$s$^{-1}$ ‡ (7,8) | $p_{on(BA)}$ | 0.1 |
| $k_{off}$ BCR-Ag | 1-$10^{-4}$ s$^{-1}$ ‡ (7,8) | $p_{off(BA)}$ | $10^{-3}$-$10^{-7}$ |
| BCR molecules/cell | ~$10^5$ (33) | $B_0$ | 500 molecules |
| Antigen concentration | 10-100 molec./μm$^2$ ‡ (7) | $A_0$ | 20-200 molecules |
| $K_A$ Ig-α/β-Lyn | $10^6$ M$^{-1}$ †‡ | $P_{A(Lyn)}$ | $10^2$ |
| $k_{on}$ Ig-α/β-Lyn | ~$10^5$ molec.$^{-1}$ s$^{-1}$ †‡ | $p_{on(Lyn)}$ | 1.0 |
| $k_{off}$ Ig-α/β-Lyn | ~10-0.1 s$^{-1}$ †‡ | $p_{off(Lyn)}$ | 0.01 |
| $K_A$ Ig-α/β-Syk | $10^6$ M$^{-1}$ (37) | $P_{A(Syk)}$ | $10^2$ |
| $k_{on}$ Ig-α/β-Syk | ~$10^5$ molec.$^{-1}$ s$^{-1}$ †‡ | $p_{on(Syk)}$ | 1.0 |
| $k_{off}$ Ig-α/β-Syk | ~10-0.1 s$^{-1}$ †‡ | $p_{off(Syk)}$ | 0.01 |
| Lyn molecules/cell | $2*10^{4}$ †‡ | $L_0$ | 100 |
| Syk molecules/cell | $4*10^{5}$ †‡ | $S_0$ | 400 |
| $k_{phos(Igα)}$ | ~100 s$^{-1}$ †‡ | $p_{phos(Igα)}$ | 0.1 |
| $k_{phos(Igβ)}$ | ~100 s$^{-1}$ †‡ | $p_{phos(Igβ)}$ | 0.1 |
| $k_{phos(Syk)}$ | ~100 s$^{-1}$ †‡ | $p_{phos(Syk)}$ | 0.1 |
| $D_{free\ molecules}$ | 0.1 μm$^2$/sec (38) | $p_{diff(F)}$ | 1.0 |
| $D_{complexes}$ | ~0.01 μm$^2$/sec (21) | $p_{diff(C)}$ | 0.1 |

† Represents a ballpark value calculated from (35,36).
‡ Parametric study performed.

FIGURE LEGENDS

**Figure 1.** Schematic of the portion of the simplified B cell receptor signaling pathway simulated in our Monte Carlo method. Antigen may bind to BCR with probability $p_{on(BA)}$. Once the same antigen molecule has stayed bound to the BCR for a threshold length of time $\mu$, Lyn may bind to either the Ig-α or Ig-β subunit with probability $p_{on(Lyn)}$ and phosphorylate both with probability $p_{phos(Ig-\alpha)}$ and $p_{phos(Ig-\beta)}$, respectively. In the meantime, Syk can freely diffuse in the cytoplasm with probability $p_{diff}$. Once the Ig-α or Ig-β subunits are phosphorylated, Syk may bind to them with probability $p_{on(Syk)}$ and become phosphorylated with probability $p_{phos(Syk)}$. Subsequent antigen binding may occur, but without any consequences.

**Figure 2.** Histogram of the numbers of bound antigen molecules. BCR-antigen affinity is varied by orders of magnitude across the physiological range in B cells, $K_A=10^5$ M$^{-1}$ to $K_A=10^{10}$ M$^{-1}$. Because of the probabilistic nature of our simulation, one hundred trials were performed for each affinity value. The parameter values used are those listed in the right hand side column of Table I, simulation time is $10^5$ time steps (corresponding to $T=100$ physical seconds).

**Figure 3.** Histogram plots for the number of BCRs with phosphorylated ITAMs (denoted as pBCR, Fig. 3A-C) and activated Syk molecules (denoted as aSyk, Fig 3D-F) after $T=100$ seconds for threshold time values of $\mu=0$ (Fig. 3A,D), $\mu=1$ second (Fig. 3B,E), $\mu=10$ seconds (Fig 3C,F). The parameter values used are those listed in the right hand side column of Table I. It only is possible to clearly distinguish between affinity values with $\mu=10$ seconds.

**Figure 4.** Plot of the mean number of bound antigen (Fig 4A), BCRs with phosphorylated ITAMs (Fig. 4B) and activated Syk molecules (Fig. 4C) as a function of affinity for threshold times $\mu=0$, $\mu=1$ second, and $\mu=10$ seconds. Where histograms are plotted in Figure 3, the mean value of each of these histograms is shown here. The number of bound antigen shows little variation with dwell time, in contrast to the number of pBCR and aSyk. A monotonic increase in signaling strength with affinity is only observed with $\mu=10$ seconds.

**Figure 5.** Plot of the number of bound antigen as a function of time for each order of magnitude in affinity between $K_A=10^5$ and $K_A=10^{10}$ M$^{-1}$.

**Figure 6.** Plot of the mean number pBCR (Fig. 6A-C) and aSyk (Fig. 6D-F) as a function of time for dwell time values $\mu=0$ (Fig. 6A,D), $\mu=1$ second (Fig. 6B,E), $\mu=10$ seconds (Fig. 6C,F). The data points for $T=100$ seconds correspond to the data points in Figure 4.

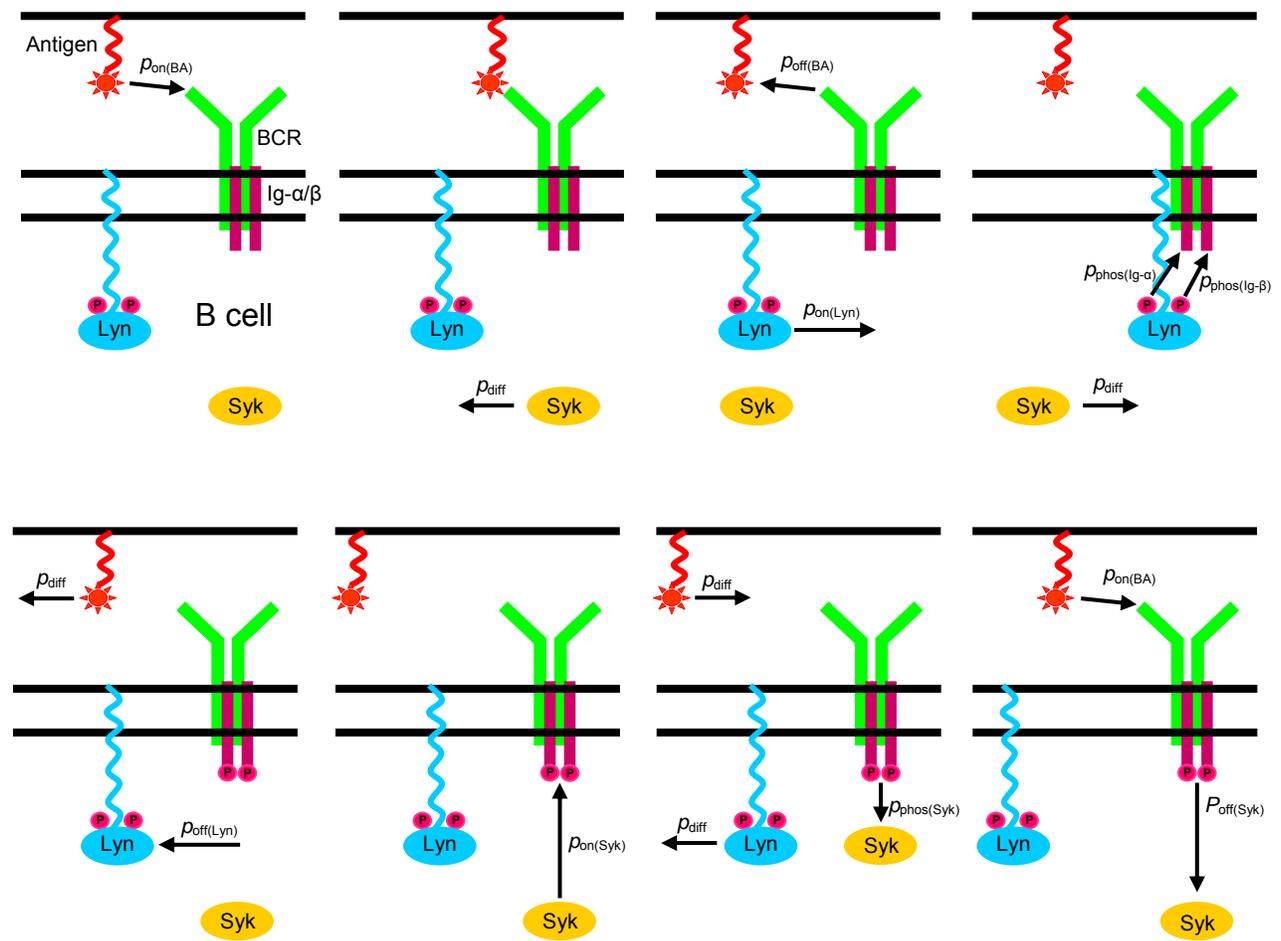

Figure 1.

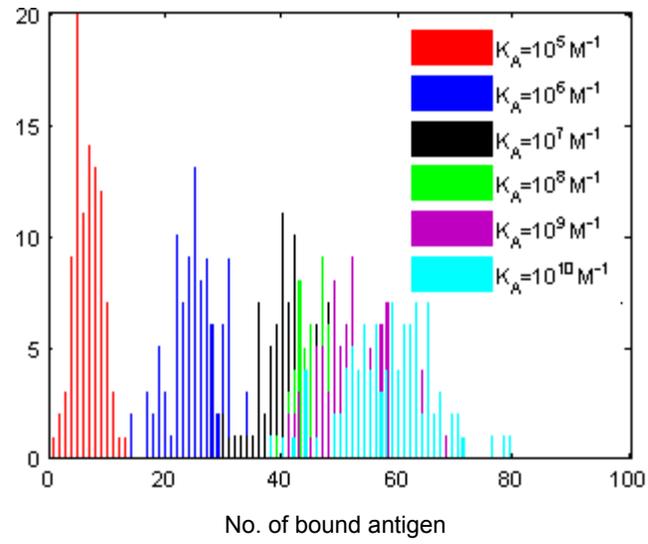

Figure 2.

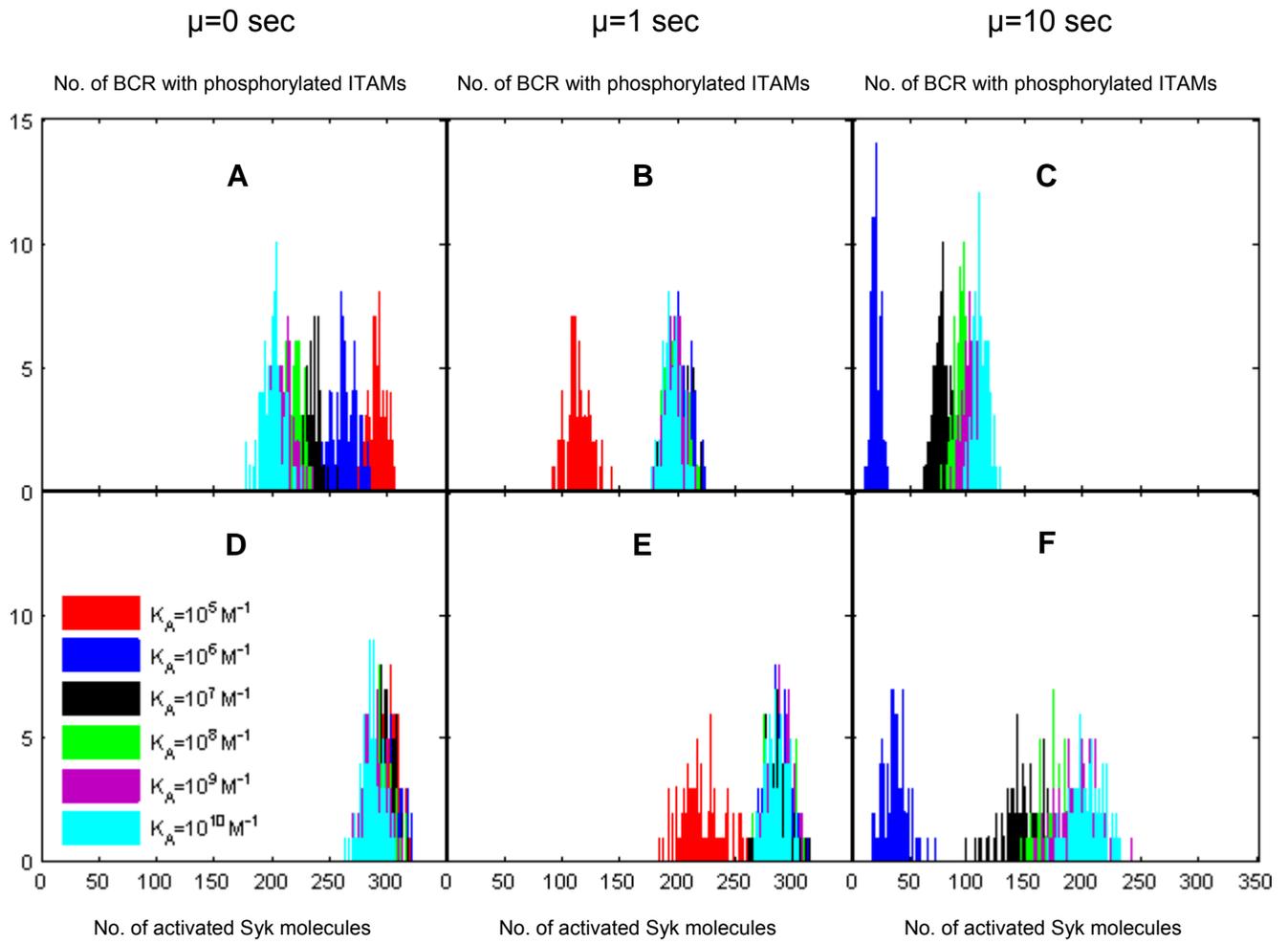

Figure 3.

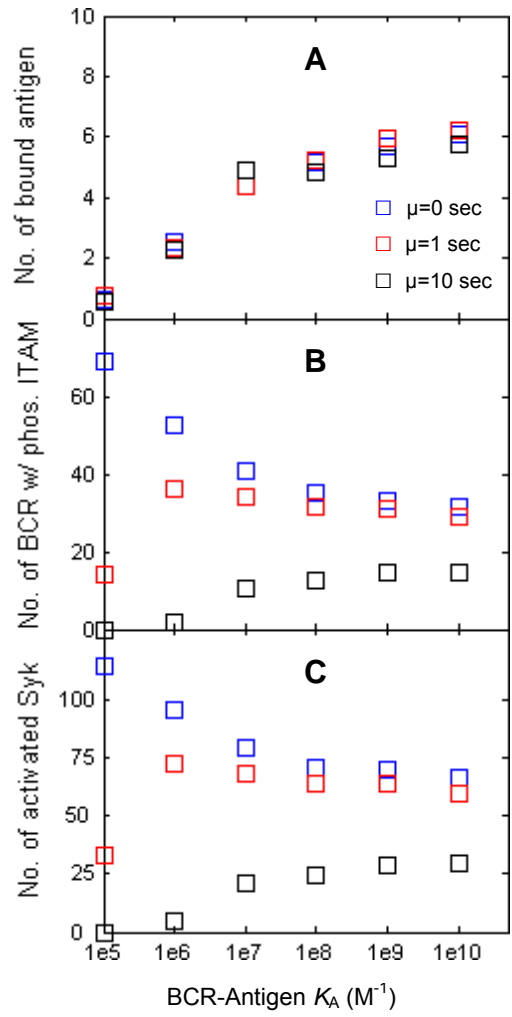

Figure 4.

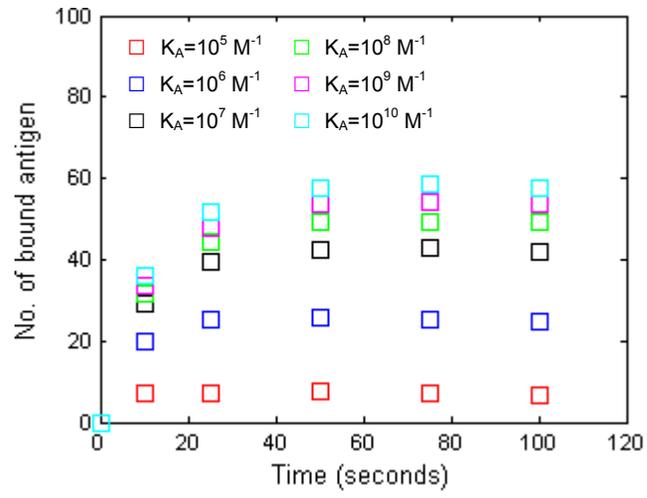

Figure 5.

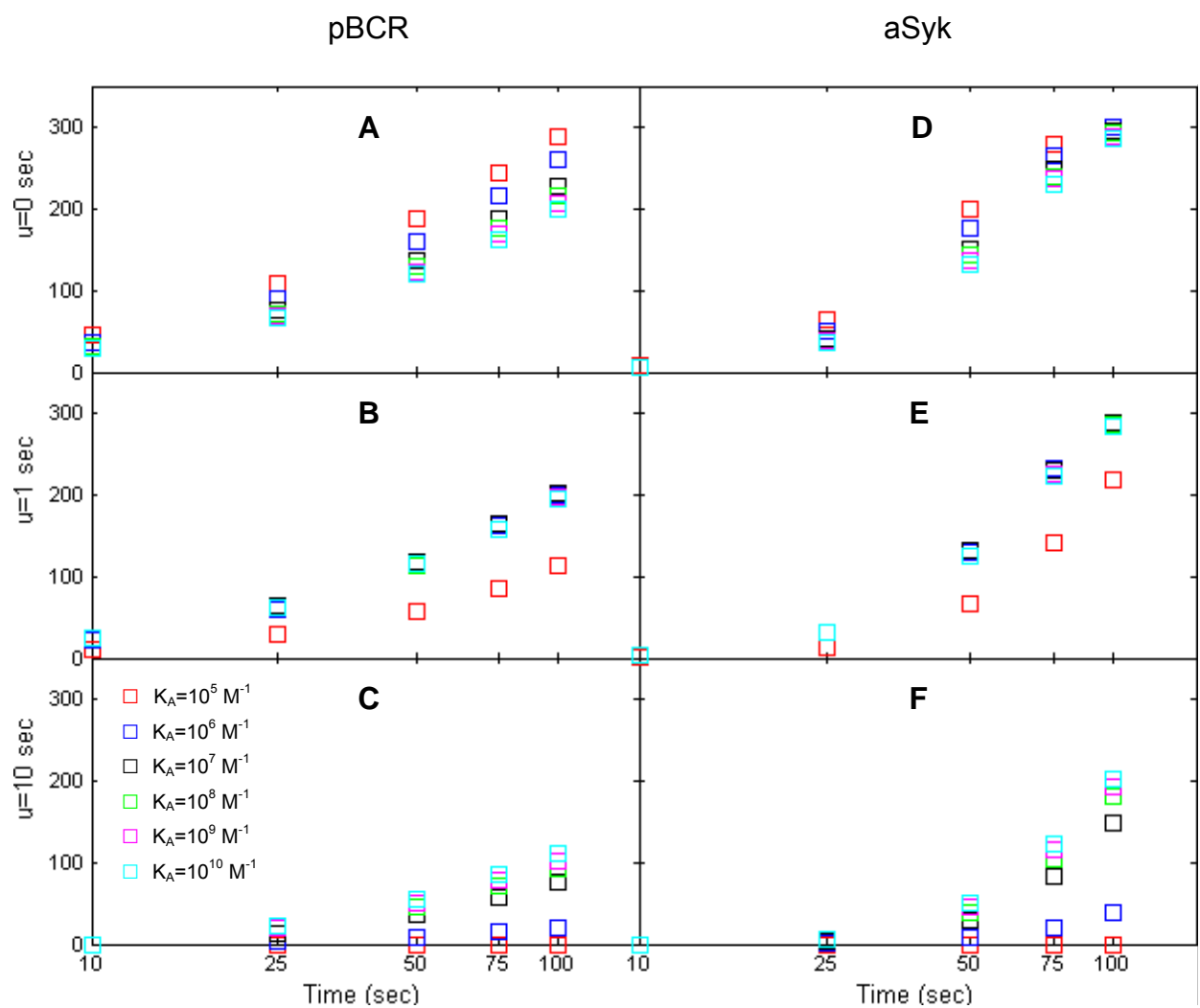

Figure 6.

SUPPLEMENTAL FIGURE LEGENDS

**Supplemental Figure 1.** Effect of varying the concentration of antigen on the mean number of bound antigen, pBCR, and aSyk molecules (100 trials). In this set of *in silico* experiments, the initial number of antigen molecules is set to $A_0$=20 molecules (compared to $A_0$=200 in the main text), which approximately corresponds to a concentration of 10 molecules/$\mu m^2$. The remaining parameter values are identical to those in Table I. The affinity discrimination pattern is identical to that of Figure 4 of the main text, even though the number of bound antigen, pBCR and aSyk molecules is different from Figure 4.

**Supplemental Figure 2.** Effect of varying the concentration of Lyn on the mean number of pBCR and aSyk molecules (100 trials). Antigen binding occurs upstream of Lyn binding, hence the number of bound antigen molecules is unaffected and not shown. In this set of *in silico* experiments, the initial number of Lyn molecules is set to $L_0$=1 molecule (compared to $L_0$=100 in the main text), with the remaining parameter values the same as Table I and $A_0$=200. We note that the affinity discrimination pattern is similar to that of Figure 4 of the main text, and even a single Lyn molecule can generate non-negligible numbers of signaling-active molecules.

**Supplemental Figure 3.** Effect of varying the concentration of Syk on the mean number of aSyk molecules (100 trials). In this set of *in silico* experiments, the initial number of Syk molecules is set to $S_0$=100 molecules, with the remaining parameter values the same as in Table I, $A_0$=200, and $L_0$=100. The affinity discrimination pattern is the same as in Figure 4 of the main text, even though the number of aSyk is different.
.

**Supplemental Figure 4.** Effect of varying Lyn kinetics on the number of pBCR and aSyk molecules. In panels A and B, the affinity of Lyn for Ig-α/β is set to $K_A$=$10^4$ $M^{-1}$, two orders below the value of $K_A$=$10^6$ $M^{-1}$ used in Table I. In panels A and B, the affinity of Lyn for Ig-α/β is set to $K_A$=$10^6$ $M^{-1}$, two orders above the value of $K_A$=$10^6$ $M^{-1}$ used in Table I. In panels E and F, the affinity of Lyn is set to $K_A$=$10^6$ $M^{-1}$ as in Table I, but the values of $p_{on(Lyn)}$ and $p_{off(Lyn)}$ are set to $P_{A(Lyn)}$=0.1/0.001, in contrast to $P_{A(Lyn)}$=1/0.01 used in the main text. In all cases, the affinity discrimination pattern is similar to that seen in Figure 4 of the main text, indicating that B cell affinity discrimination is largely independent of Lyn kinetics.
.

**Supplemental Figure 5.** Effect of varying Syk kinetics on the number of aSyk molecules. The affinity of Syk is set to the literature value of $K_A$=$10^6$ $M^{-1}$ used in Table I, but the values of $p_{on(Syk)}$ and $p_{off(Syk)}$ are set to $P_{A(Syk)}$=0.1/0.001, in contrast to $P_{A(Syk)}$=1/0.01 used in the main text. The affinity discrimination pattern observed is identical to that of Figure 4 of the main text, indicating that B cell affinity discrimination is not dependent on Syk kinetics.

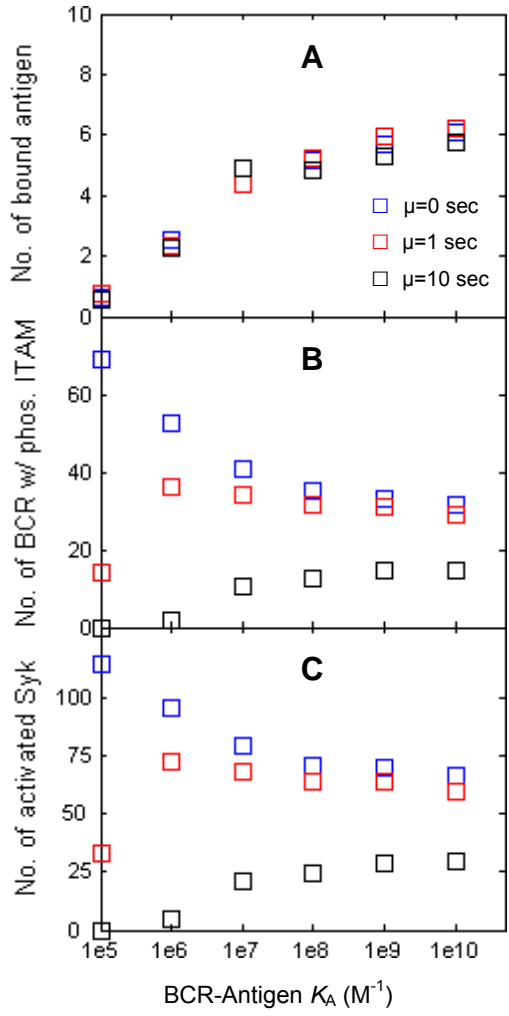

Figure S.1.

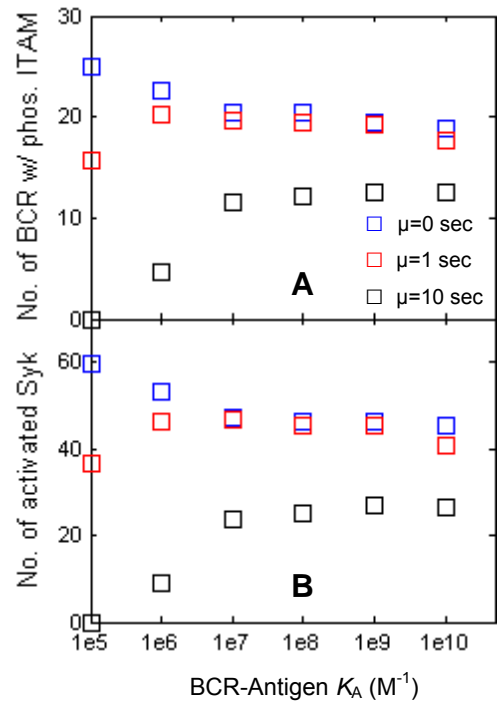

Figure S.2.

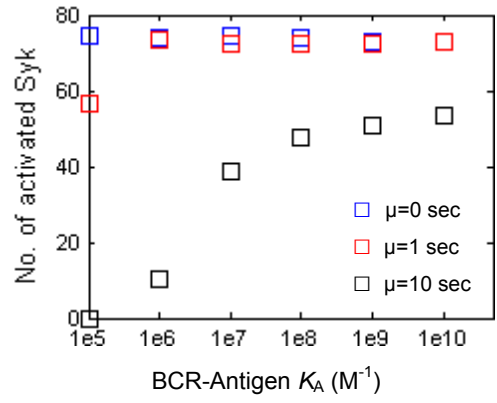

Figure S.3.

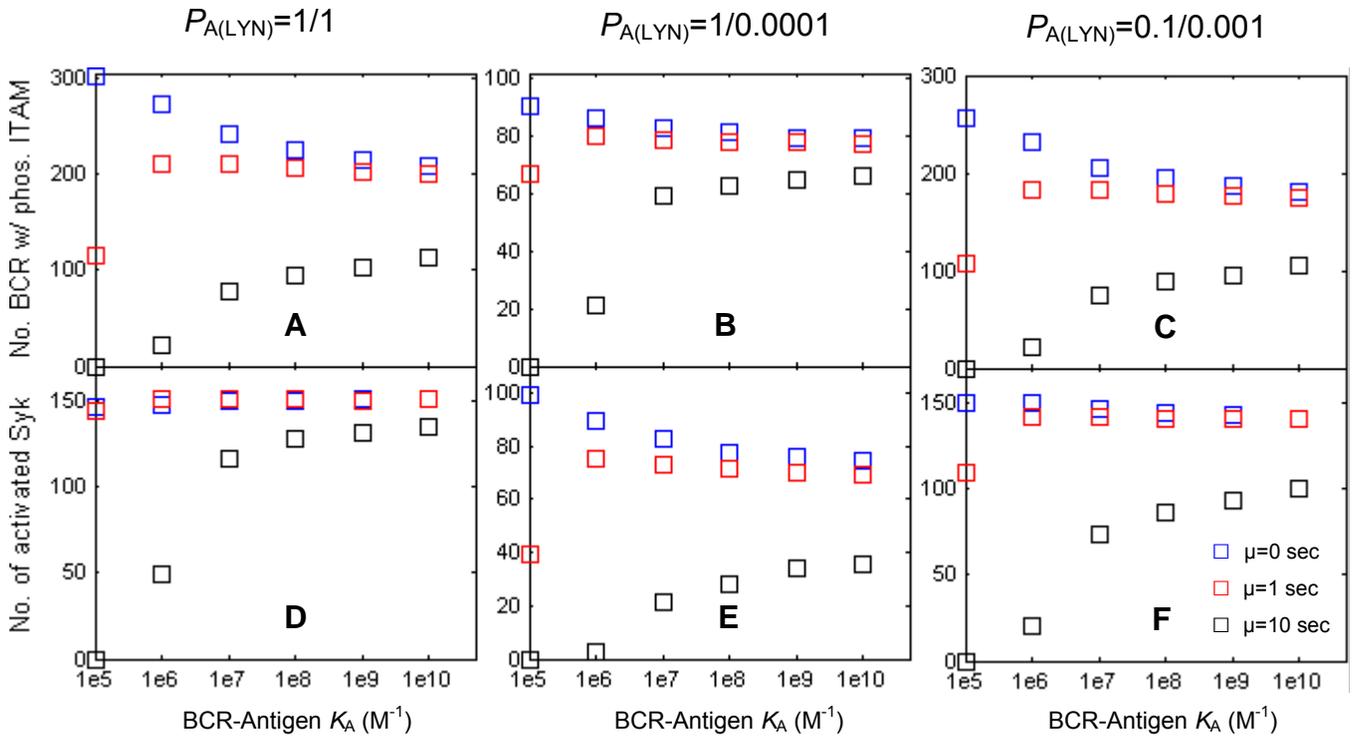

Figure S.4.

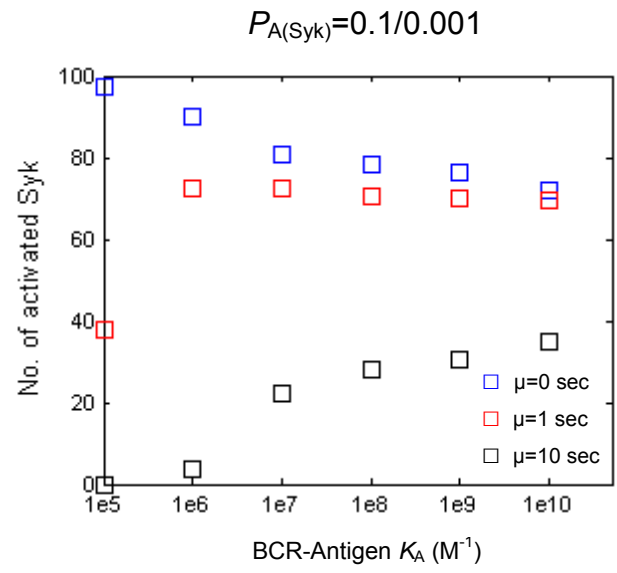

Figure S.5.